\documentclass[aps,twocolumn,tightenlines]{revtex4}
\usepackage[dvipdf]{graphicx}

\begin{document}
\title{Atom interferometry measurement of the electric polarizability of lithium}

\author{A. Miffre, M. Jacquey, M. B\"uchner, G. Tr\'enec and J. Vigu\'e}
\address{ Laboratoire Collisions Agr\'egats R\'eactivit\'e -IRSAMC
\\Universit\'e Paul Sabatier and CNRS UMR 5589
 118, Route de Narbonne 31062 Toulouse Cedex, France
\\ e-mail:~{\tt jacques.vigue@irsamc.ups-tlse.fr}}

\date{\today}

\begin{abstract}

Using an atom interferometer, we have measured the static electric
polarizability of $^7$Li  $\alpha =(24.33 \pm 0.16)\times10^{-30}
$ m$^3$ $= 164.19\pm 1.08 $ atomic units with a $0.66$\%
uncertainty. Our experiment, which is similar to an experiment
done on sodium in 1995 by D. Pritchard and co-workers, consists in
applying an electric field on one of the two interfering beams and
measuring the resulting phase-shift. With respect to D.
Pritchard's experiment, we have made several improvements which
are described in detail in this paper: the capacitor design is
such that the electric field can be calculated analytically; the
phase sensitivity of our interferometer is substantially better,
near $16$ mrad/$\sqrt{\mbox{Hz}}$; finally our interferometer is
species selective so that impurities present in our atomic beam
(other alkali atoms or lithium dimers) do not perturb our
measurement. The extreme sensitivity of atom interferometry is
well illustrated by our experiment: our measurement amounts to
measuring a slight increase $\Delta v$ of the atom velocity $v$
when it enters the electric field region and our present
sensitivity is sufficient to detect a variation $\Delta v/v
\approx 6 \times 10^{-13}$.

\end{abstract}
\maketitle


\section{Introduction}

The measurement of the electric polarizability $\alpha$ of an atom
is a difficult experiment: this quantity cannot be measured by
spectroscopy, which can access only to polarizability differences,
and one should rely either on macroscopic quantity measurements
such as the electric permittivity (or the index of refraction) or
on electric deflection of an atomic beam. For a review on
polarizability measurements, we refer the reader to the book by
Kresin and Bonin \cite{bonin97}. For alkali atoms, all the
accurate experiments were based on the deflection of an atomic
beam by an inhomogeneous electric field and, in the case of
lithium, the most accurate previous measurement was done in 1974
by Bederson and co-workers \cite{molof74}, with the following
result $\alpha = (24.3 \pm 0.5)\times10^{-30}$ m$^3$.  However, in
2003, Amini and Gould, using an atomic fountain \cite{amini03},
have measured the polarizability of cesium atom with a $0.14$\%
relative uncertainty, which is presently the smallest uncertainty
on the electric polarisability of an alkali atom.

Atom interferometry, which can measure any weak modification of
the atom propagation, is perfectly adapted to measure the electric
polarizability of an atom: this was demonstrated in 1995 by D.
Pritchard and co-workers \cite{ekstrom95} with an experiment on
sodium atom and they obtained a very high accuracy, with a
statistical and systematic uncertainties both equal to $0.25$\%.
This experiment was and remains difficult because an electric
field must be applied on only one of the two interfering beams:
one must use a capacitor with a thin electrode, a septum, which
can be inserted between the two atomic beams.

Using our lithium atom interferometer
\cite{delhuille02a,miffre05}, we have made an experiment very
similar to the one of D. Pritchard \cite{ekstrom95} and we have
measured the electric polarizability of lithium with a $0.66$\%
uncertainty, limited by the uncertainty on the mean atom velocity
and not by the atom interferometric measurement itself
\cite{miffre05a}. In the present paper, we are going to describe
in detail our experiment with emphasis on the improvements with
respect to the experiments of D. Pritchard's group
\cite{ekstrom95,roberts04}: we have designed a capacitor with an
analytically calculable electric field;  we have obtained a
considerably larger phase sensitivity, thanks to a large atomic
flux and an excellent fringe visibility; finally our
interferometer, which uses laser diffraction, is species
selective: the contribution of any impurity (heavier alkali atoms,
lithium dimers) to the signal can be neglected.

We may recall that several experiments using atom interferometers
have exhibited a sensitivity to an applied electric field
\cite{shimizu92,nowak98,nowak99} but these experiments were not
aimed at an accurate measurement of the electric polarizability.
Two other atom interferometry experiments
\cite{rieger93,morinaga96} using an inelastic diffraction process,
so that the two interfering beams are not in the same internal
state, have measured the difference of polarizability between
these two internal states. Finally, two experiments
\cite{sangster93,sangster95,zeiske95} have measured the
Aharonov-Casher phase \cite{aharonov84}: this phase, which results
from the application of an electric field on an atom with an
oriented magnetic moment, is proportional to the electric field.

This paper is organized as follows. We briefly recall the
principle of the experiment in part II. We then describe our
electric capacitor in part III and the experiment in part IV. The
analysis of the experimental data is done in part V and we discuss
the polarizability result in part VI. A conclusion and two
appendices complete the paper.

\section{Principle of the measurement}

If we apply an electric field $E$ on an atom, the energy of its
ground state decreases by the polarizability term:

\begin{equation}
\label{n1} U = - 2 \pi \epsilon_0 \alpha E^2
\end{equation}

When an atom enters a region with a non vanishing electric field,
its kinetic energy increases by $-U$ and its wave vector $k$
becomes $ k +\Delta k$, with $\Delta k$ given by $\Delta k = 2 \pi
\epsilon_0 \alpha E^2 m/(\hbar k)$. The resulting phase shift
$\phi $ of the atomic wave is given by:

\begin{equation}
\label{n2} \phi = \frac{2 \pi \epsilon_0 \alpha }{\hbar v} \int
E^2(z) dz
\end{equation}

\noindent where we have introduced the atom velocity $v = \hbar
k/m$ and taken into account the spatial dependence of the electric
field along the atomic path following the $z$-axis. This phase
shift is inversely proportional to the atom velocity and this
dependence will be included in our analysis of the results.

\begin{figure}
\includegraphics[width = 7 cm,height= 5 cm]{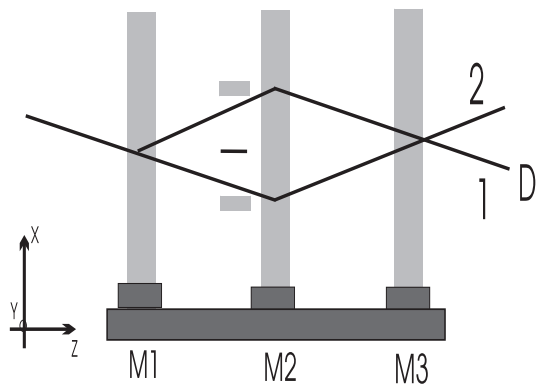}
\caption{\label{interferometer} Schematic drawing of our
experiment based on a Mach-Zehnder atom interferometer: a
collimated atomic beam, coming from the left, is diffracted by
three laser standing waves made by reflecting three laser beams on
the mirrors $M_1$, $M_2$ and $M_3$. The output beam labelled $1$
is selected by a slit and detected by a hot-wire detector D. The
capacitor with a septum between the two interfering beams is
placed just before the second laser standing wave. The $x$, $y$
and $z$ axis are defined.}
\end{figure}

The principle of the experiment, illustrated on figure
\ref{interferometer}, is to measure this phase shift by applying
an electric field on one of the two interfering beams in an atom
interferometer \cite{ekstrom95}. This is possible only if the two
beams are spatially separated so that a septum can be inserted
between the two beams. This requirement could be suppressed by
using an electric field with a gradient as in reference
\cite{roberts04} but it seems difficult to use this arrangement
for a high accuracy measurement, because an accurate knowledge of
the values of the field and of its gradient at the location of the
atomic beams would be needed.

\section{The electric capacitor}

\subsection{Capacitor design}

\begin{figure}
\includegraphics[width = 7 cm,height= 5 cm]{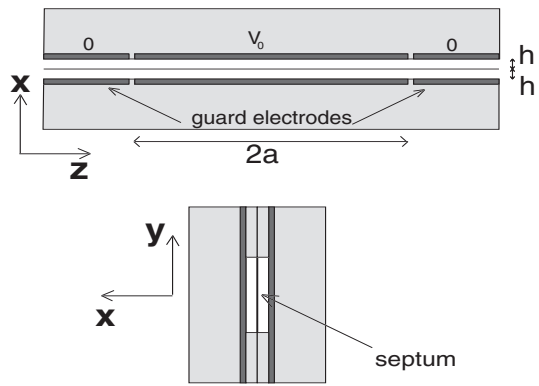}
\caption{\label{capacitor} Two schematic drawings of the
capacitor: top view of the capacitor cut in the atomic beam plane
(upper panel) and end view as seen by an observer located on the
atomic beam (lower panel). The axis being the same as in figure 1,
the septum and the electrodes are parallel to the $y$, $z$ plane,
with the septum at $x=0$ and the electrodes at $x= \pm h$. The
high voltage electrodes at the $V=V_0$ potential extends from
$z=-a$ to $z=+a$, while the guard electrodes extend outside with
$|z|>a$. The septum and the guard electrodes are at the $V=0$
potential.}
\end{figure}

\begin{figure}
\includegraphics[width = 7 cm,height= 6 cm]{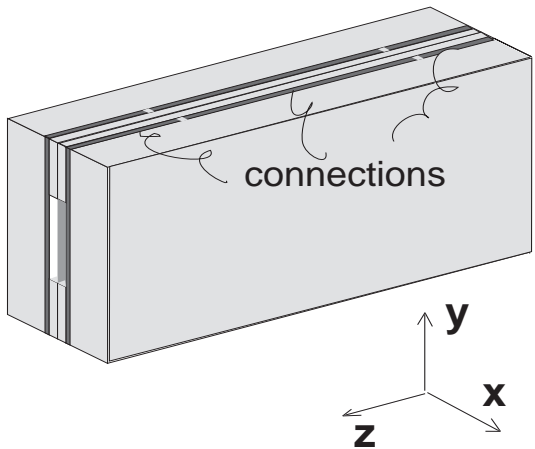}
\caption{\label{artistview} Artist's view of the capacitor: we
have also shown schematically some wires connecting the
electrodes, the important point being that these wires are not
close to the atomic beams.}
\end{figure}

To make an accurate measurement, we must know precisely the
electric field along the atomic path and guard electrodes are
needed so that the length of the capacitor is well defined, as
discussed by D. Pritchard and co-workers \cite{ekstrom95}. It
would probably be better to have guard electrodes on both
electrodes, but it seems very difficult to draw guard electrodes
on the septum and to put them in place very accurately. Therefore,
as in reference \cite{ekstrom95}, we have guard electrodes only on
the massive electrodes. However, we have chosen to put our guard
electrodes in the plane of the high voltage electrode. With this
choice, the calculation of the electric field can be done
analytically. Figure \ref{capacitor} presents two schematic
drawings of the capacitor and defines our notations, while an
artist's view is presented in figure \ref{artistview}. Like in
reference \cite{ekstrom95}, our capacitor is as symmetric as
possible with respect to the septum plane, but, for a given
experiment, only one half of the capacitor is used, the other part
creating no electric field with $V=0$ everywhere.


\subsection{Calculation of the capacitor electric field}

\subsubsection{From three dimensions to two dimensions}

If a dielectric slab with a permittivity $\epsilon_r
> 1$ is introduced in a plane capacitor, the field lines are
distorted and concentrated towards the slab. Because our capacitor
contains dielectric spacers, one could fear a similar effect but
this effect does not exist when the dielectric slab fills
completely the gap between the electrodes. Following figure
\ref{capacitor}, the spacers, with a dielectric constant
$\varepsilon_r$ completely fill the space $\left|y\right| > y_0$,
with vacuum in the rest of capacitor, $\left|y\right| < y_0$. Let
$V$ (respectively $V_1$) be the potential with $\left|y\right| <
y_0$ ($\left|y\right| > y_0$). In the $\left|y\right| = y_0$
planes, the continuity equations give:

\begin{equation}
\label{n3} V = V_1 {\mbox{   and  }} \frac{\partial V}{\partial y}
= \frac{\partial V_1}{\partial y}
\end{equation}

\noindent As shown below, a {\bf 2D} solution of the Laplace
equation of the form $V(x,z)$ exists for the capacitor. Then
clearly, if we take $V_1= V$, this solution fulfills both
continuity equations on the dielectric borders in the
$\left|y\right| = y_0$ planes and we know that the solution is
unique.

\subsubsection{Calculation of the potential and of the electric field}

We consider only one half (with $x\geq 0$) of the capacitor
represented on figure \ref{capacitor}. We know the potential on
the borders of the capacitor, with $V(x=0,z) = 0$ and $V(x= h, z)
= V_0$ if $\left|z\right| < a $ and $V(x= h, z) = 0$ if
$\left|z\right| > a $. To get the potential everywhere, we start
by calculating the Fourier transform $\tilde{V}(k)$ of $V(x= h,
z)$ :

\begin{equation}
\label{n4} \tilde{V}(k)  =  \frac{1}{\sqrt{2\pi}}
\int_{-\infty}^{+\infty}V(x= h, z) \exp\left( -ikz\right) dz
\end{equation}

\noindent Then, for a separable harmonic function of $x$ and $z$,
with a $z$-dependence of the $\exp\left(ikz\right)$ form, its
$x$-dependence is necessarily given by a linear combination of the
two functions $\exp\left(\pm kx\right)$. Using this result and the
conditions on the borders at $x=0$ and at $x=h$, we get the value
of $V(x,z)$ everywhere:

\begin{equation}
\label{n5} V(x,z) =  \frac{1}{\sqrt{2\pi}}
\int_{-\infty}^{+\infty} \tilde{V}(k) \frac{\sinh\left(k
x\right)}{\sinh\left(k h\right)} \exp\left( ikz\right) dk
\end{equation}

\noindent from which we can deduce the electric field everywhere.
On the septum surface $x= 0$, the electric field is parallel to
the $x$-axis and we can calculate exactly the integral of $E^2$ on
this surface (see appendix A).  We need the capacitor effective
length $L_{eff}$ which is defined by:

\begin{equation}
\label{n9} L_{eff}= \frac{1}{E_0^2} \int_{-\infty} ^{+\infty} E^2
dz
\end{equation}

\noindent where $E_0 = V_0/h$ is the electric field of an infinite
plane capacitor with the same electrode spacing $h$. Using
equation (\ref{a51}), we get the exact value of $L_{eff}$:

\begin{equation}
\label{n91} L_{eff} =  2a \left[\coth\left(\frac{\pi a}{h}\right)
- \frac{h} {\pi a}\right]  \approx  2a - \frac{2h}{\pi}
\end{equation}

\noindent where exponentially small corrections of the order of
$\exp\left[ -2 \pi a/ h \right]$ have been neglected in the
approximate result.

However, the atoms do not sample the electric field on the septum
surface but at a small distance, of the order of $50$ $\mu$m in
our experiment, and what we need is the integral of $E^2$ along
their mean trajectory. This mean trajectory is not exactly
parallel to the septum, but it is easier to calculate the integral
along a constant $x$ line. We may use either the potential given
by equation (\ref{n5}) or Maxwell's equations to relate the field
components near the septum surface to its value on the surface.
The calculation, given in appendix A, proves that the first
correction to the effective length is proportional to $x^2$:

\begin{equation}
\label{n13} L_{eff} \approx 2a - \frac{2h}{\pi} + \frac{2\pi
x^2}{3 h}
\end{equation}

\noindent  The $x^2$ correction is a fraction of the main term
$2a$ equal to $ \pi x^2/(3 a h)$  and with our dimensions ($x
\approx 50$ $\mu$m, $h \approx 2$ mm and $a  \approx 25$ mm), this
correction is close to $5 \times 10^{-5} L_{eff}$. This correction
is negligible at the present level of accuracy and we will use the
value of $L_{eff}$ given by the approximate form of equation
(\ref{n91}). More precisely, we will write:

\begin{equation}
\label{n14} \int_{-\infty} ^{+\infty} E^2 dz = V_0^2 \left[
\frac{2a}{h^2} - \frac{2}{\pi h}\right]
\end{equation}

\subsection{Construction of the capacitor}

Let us describe how we build this capacitor. The external
electrodes are made of glass plates ($80$ mm long in the $z$
direction, $35$ mm high in the $y$ direction, $10$ mm thick in the
$x$ direction) covered by an evaporated aluminium layer. To
separate the guard electrodes from the high voltage electrode, a
gap is made in the aluminium layer by laser evaporation
\cite{cheval}. We found that $100$ $\mu$m wide gaps give a
sufficient insulation under vacuum to operate the capacitor up to
$V_0 = 500$ V. These gaps are separated by a distance $2a= 50$ mm,
so that the two guard electrodes are $15$ mm long. The glass
spacers are $\sim 2$ mm thick plates of float glass ($10$ mm
$\times$ $80$ mm) used without further polishing. The distance
$y_0$ from the spacer inner edge to the atomic beam axis is equal
to $\sim 7$ mm.

We have found that a float glass plate is flat within $\pm 2 $
$\mu$m over the needed surface. This accuracy appeared to be
sufficient for a first construction, as the main geometrical
defects are due to the way we assemble the various parts by gluing
them together and by an imperfect stretching of the septum.  We
use a double-faced tape ARCLAD 7418 (from Adhesive Research) to
assemble the spacers on the external electrodes.

The septum is made of $6$ $\mu$m thick mylar from Goodfellow
covered with aluminium on both faces. In a first step, the mylar
sheet is glued on circular metal support. It is then covered by a
thin layer of dish soap diluted in water and the mylar is heated
near $80^{\circ}$C with a hot air gun. Then, we clean the mylar
surface with water and let it dry. After this operation, the mylar
is well stretched and its surface is very flat. We have measured
the resonance frequencies of the drum thus formed, from which we
deduced a surface tension of the order of $50$ N/m (this value is
only indicative as this experiment was made with another mylar
film which was $20$ $\mu$m thick). Once stretched, the mylar film
is glued on one electrode-spacer assembly with an epoxy glue
EPOTEK 301 (from Epoxy Technology), chosen for its very low
viscosity, and then it is cut with a scalpel. In a final step, a
second electrode-spacer assembly is glued on the other face of the
mylar. Finally, as shown in figure \ref{artistview}, wires are
connected to the various electrodes using an electrically
conductive adhesive EPOTEK EE129-4 (also from Epoxy Technology).

\subsection{Residual defects of the capacitor}

We are going to discuss the various points by which the real
capacitor differs from our model.

\subsubsection{2D character of the potential}

We have shown that the potential $V$ is reduced to a {\bf 2D}
function when an homogeneous dielectric slab fills completely the
gap between the electrodes, with the border between the vacuum and
the slab being a plane perpendicular to the electrodes. The real
dielectric slab is the superposition of a tape, a glass spacer and
a glue film, each material having a different permittivity
$\epsilon_r$. The differences in permittivity perturb the
potential which should take a {\bf 3D} character extending on a
distance comparable to the tape or glue film thicknesses. This
perturbation seems negligible, because the tape and the glue films
are very thin and also because these three dielectric materials
have not very large $\epsilon_r$ values.

\subsubsection{Do we know the potential everywhere on the border?}

Our calculation assumes that the potential is known everywhere on
the border. But, on the high-voltage electrode, we may fear that
the potential is not well defined in the $100$ $\mu$m wide
dielectric gaps separating the high voltage and guard electrodes
as these gaps might get charged in an uncontrolled way. This is
not likely if the volume resistivity $\rho$ of the pyrex glass
used is not too large: more precisely, the time constant for the
charge equilibration on the gap surface is given by $\epsilon_0
\rho$ (within numerical factors of the order of one) and this time
constant remains below $1$ second if $\rho \leq 10^{11}$
$\Omega$.m. We have found several values of the resistivity of
pyrex glass at ordinary temperature, in the range $(4-8)\times
10^8$ $\Omega$.m and with such a conductivity, this time constant
is below $10^{-2}$ s. Our calculation neglects the surface
conductivity, due to the adsorbed impurities, which should further
reduce this time constant.

Therefore, we think that it is an excellent approximation to
assume that the potential $V$ makes a smooth transition from $V=0$
to $V=V_0$ in the gaps. Then, using equation (\ref{a3}), it is
clear that the detailed shape of the transition has no consequence
as these details are smoothed out by the convolution of $V(z)$ by
the function $g(z)$ which has a full width at half maximum equal
to $1.12 h $. We can use equation (\ref{n91}) to calculate the
effective length, provided that we add to the length of the high
voltage electrode the mean width of the two gaps. In the present
work, we have taken one gap width, $100$ $\mu$m, as a conservative
error bar on the effective length. A superiority of our capacitor
design is that these gaps are very narrow, thus minimizing the
corresponding uncertainty on the capacitor effective length and we
hope to be able to further reduce this uncertainty.

\subsubsection{Parallelism of the electrodes}

The thickness $h$ of the capacitor, which is the sum of the
thickness of the spacers, the tape and the glue film, is not
perfectly constant. Using a Mitutoyo Litematic machine, we have
measured, with a $\pm 1$ $\mu$m uncertainty, the capacitor
thickness as a function of $z$, in the center line of the two
spacers, at $y= \pm 12$ mm. The average of these two measurements
gives the thickness $h(z)$ in the $y=0$ plane around which the
atom sample the electric field. The thickness $h$ is not perfectly
constant but it is well represented by a linear function of $z$,
given by $ h(z) = h_0 + h_1(z/a)$, the maximum deviation $h_1$
being considerably smaller than the mean $h$ value noted $h_0$. As
these deviations are very small (see below), it seems reasonable
to use use equation (\ref{n14}) provided that terms involving
powers of $h$ are replaced by their correct averages. The first
term in $h^{-2}$ corresponds to the integral of $E^2$ over the
capacitor length, from $z=-a$ to $z= a$ and we must take the
average value of $h^{-2}$ over this region. Neglecting higher
order terms, this average is given by:
\begin{equation}
\label{n15} \left<\frac{1}{h^2}\right> = \frac{1}{h_0^2} \left[1 +
\frac{h_1^2}{h_0^2}\right]
\end{equation}

\noindent In equation (\ref{n14}), the second term in $h^{-1}$,
corresponds to end effects and this quantity must be replaced by
the following two-point average:

\begin{equation}
\label{n16} \left<\frac{1}{h}\right> =
\frac{1}{2}\left(\frac{1}{h(z=-a)} + \frac{1}{h(z=a)}\right) =
\frac{1}{h_0}  \left[1 + \frac{h_1^2}{h_0^2}\right]
\end{equation}
\noindent  Both corrections involve the same factor $[1 +
(h_1^2/h_0^2)]$.

\subsubsection{Summary of the capacitor dimensions}

Although the capacitor is as symmetric as possible, this symmetry
is only approximate and we give the parameters for the half we
have used for the set of measurements described below. The length
of the high voltage electrode, including one gap width is $2a =
50.00 \pm 0.10$ mm, the error bar being taken equal to one gap
width, as discussed above. The distance $h$ between the electrodes
gap width is described by $h_0 = 2.056 \pm 0.003 $ mm and $h_1 =
3.2\times 10^{-3}$ mm. The correction term $h_1^2/h_0^2 = 2.4
\times 10^{-6}$ is completely negligible.

\section{The experiment}

In this part, we are going to recall the main features of our
lithium atom interferometer, to give the values of various
parameters used for the present study, to present the data
acquisition procedure and the way we extract the phases from the
data.

\subsection{Our interferometer}

Our atom interferometer is a Mach-Zehnder interferometer using
Bragg diffraction on laser standing waves. Its design is inspired
by the sodium interferometer of D. Pritchard and co-workers
\cite{keith91,schmiedmayer97} and by the metastable neon
interferometer of Siu Au Lee and co-workers \cite{giltner95b}. A
complete description has been published
\cite{delhuille02a,miffre05}.

The lithium atomic beam is a supersonic beam seeded in argon and,
for the present experiment, we have worked with a low argon
pressure in the oven $p_0 = 167$ mbar, because the detected
lithium signal increases when the argon pressure decreases. The
oven body temperature is equal to $973$ K, fixing the vapor
pressure of lithium at $p_{Li} = 0.8$ mbar and the nozzle
temperature is equal to $T_0= 1073$ K. With these source
conditions, following our detailed analysis \cite{miffre05b}, the
argon and lithium velocity distributions are described by a
parallel speed ratio for argon equal to $S_{\|,Ar} = 8.3$ and a
parallel speed ratio for lithium equal to $S_{\|,Li} = 6.2$ (the
parallel speed ratio is defined by equation (\ref{m4}) below).

We use Bragg diffraction on laser standing waves at $\lambda =
671$ nm: the laser is detuned by about $3$ GHz on the blue side of
the $^2S_{1/2}$ - $^2P_{3/2}$ transition of the $^7$Li isotope,
the signal is almost purely due to this isotope, which has a
natural abundance equal to $92.41$\%, and not to the other isotope
$^6Li$. Moreover, any other species present in the beam, for
instance heavier alkali atoms or lithium dimers, is not diffracted
and does not contribute to the signal.

The case of lithium dimers deserves a special discussion because
they are surely present in the beam and the lithium dimer has an
absorption band system due its $A-X$ transition with many lines
around $671$ nm: for most rovibrational levels of the $X$ state,
the absorption transition which is closest to the laser frequency
has a very large detuning of the order of hundreds to thousands of
GHz and the intensity of this resonance transition is also weaker
than the resonance transition of lithium atom, because of the
Franck-Condon factor. Therefore, the lithium dimers have a
negligible probability of diffraction and do not contribute to the
interferometer signals.

The interference signals in a Mach-Zehnder interferometer are
given by:

\begin{equation}
\label{ex1} I = I_{0} \left[1 + {\mathcal{V}}\cos \psi \right]
\end{equation}

\noindent where the phase $\psi$ of the interference fringes can
be written:
\begin{equation}
\label{ex2} \psi =  p (2k_L) (x_1+x_3-2x_2) + \phi
\end{equation}

\noindent The first term  of $\psi$ is particular to three-grating
interferometers: the diffracted beam of order $p$ by grating $i$
has a phase dependent on the grating position $x_i$. In our case
of laser diffraction, the grating position $x_i$ is given by the
mirror position $M_i$ and $2k_L$ is the grating wavevector, where
$k_L$ is the laser wavevector. This phase term is very interesting
because it is non dispersive and it is commonly used to observe
interference fringes. In our case, we scan the position $x_3$ of
mirror $M_3$ by a piezoelectric translation stage. The second term
$\phi $ represent any phase difference between the two beams and
in particular, it will represent the phase shift due to the
application of an electric field on one of the two paths.

\subsection{Introduction of the capacitor}

\begin{figure}
\includegraphics[width = 8 cm,height= 6 cm]{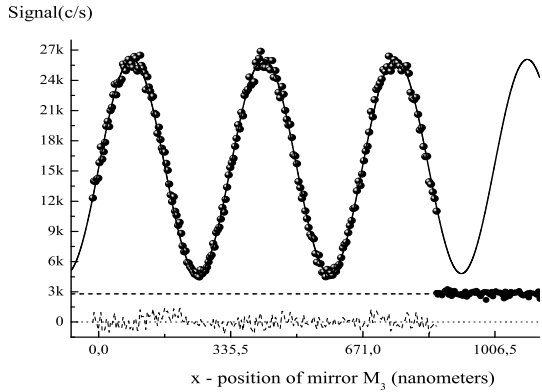}
\caption{\label{fringes84} Interference signal observed by
displacing the mirror $M_3$: the signal is expressed in counts per
second, but the counting time is equal to $0.1$ second only and
the background signal has been recorded at the end of the
experiment. The full curve is the best fit and the fit residuals
are plotted, at the bottom of the figure.}
\end{figure}

In the present work, we have only used the diffraction order $p=1$
so that the center of the two beams are separated by about $90$
$\mu$m at the location of the capacitor, which is located just
before the second laser standing wave. The capacitor is attached
to the top of the vacuum chamber and not to the rail supporting
the laser standing wave mirrors $M_i$: in this way, we do not
increase the vibrations of the mirror positions $x_i$ The
capacitor is held by a translation stage along the $x$-direction,
which can be adjusted manually thanks to a vacuum feedthrough and
a double stage kinematic mount built in our laboratory. The first
stage, operated with screws, can be used only when the experiment
is at atmospheric pressure while the second stage, actuated by
low-voltage piezo-translators, can be adjusted under vacuum. When
the septum is inserted between the two atomic paths, the atom
propagation is almost not affected by its presence and, as shown
in figure \ref{fringes84}, we have observed a fringe visibility
equal $\mathcal{V} = 84 \pm 1$ \% and a negligible reduction of
the atomic flux.

\begin{figure}
\includegraphics[width = 8 cm,height= 7 cm]{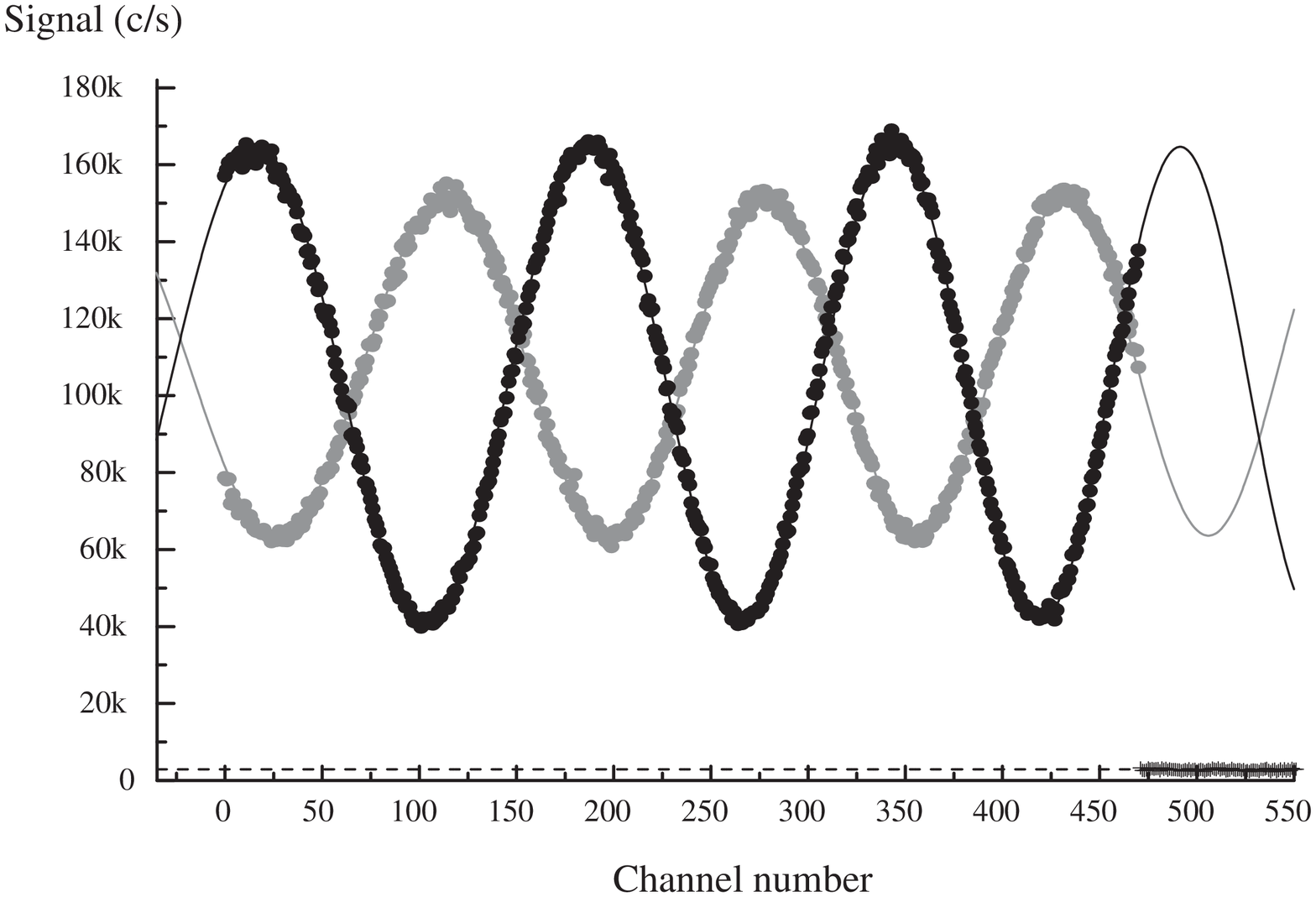}
\caption{\label{signals} Experimental signals corresponding to
$V_0=0$ (black dots) and to $V_0 \approx 260$ Volts (grey dots).
Their sinusoidal fits using equation (\ref{ex3}) are represented
by the full curves. The phase shift due to the polarizability
effect is close to $3\pi$ and the fringe visibility is reduced
because of the dispersion of the phase shift associated to the
velocity spread of the lithium atoms.}
\end{figure}

To optimize the phase sensitivity (see a discussion in reference
\cite{miffre05}), we have opened the collimation slit $S_1$ and
the detection slit $S_D$ (see reference \cite{miffre05}) with
widths $e_1= 18$ $\mu$m and $e_D = 50$ $\mu$m, thus increasing the
mean flux up to $10^5$ counts/s and slightly reducing the fringe
visibility down to ${\mathcal{V}_0} = 62$\% (see figure
\ref{signals}).

\subsection{The data acquisition procedure}

We have made a series of recordings, labelled by an index $i$ from
$1$ to $44$, with $V_0=0$ when $i$ is odd, and with $V_0 \neq 0$,
when $i$ is even with $V_0 \approx 10\times i $ Volts. For each
recording, we apply the same linear ramp on the piezo-drive of
mirror $M_3$ in order to observe interference fringes and $471$
data points are recorded with a counting time per channel equal to
$0.36$ s. Figure \ref{signals} presents a pair of consecutive
recordings.

The high voltage power supply has stability close to $ 10^{-4}$
and the applied voltage is measured by a HP model $34401A$
voltmeter with a relative accuracy better than $10^{-5}$.

\subsection{Extracting phases from the data}

\begin{figure}
\includegraphics[width = 8 cm,height= 6 cm]{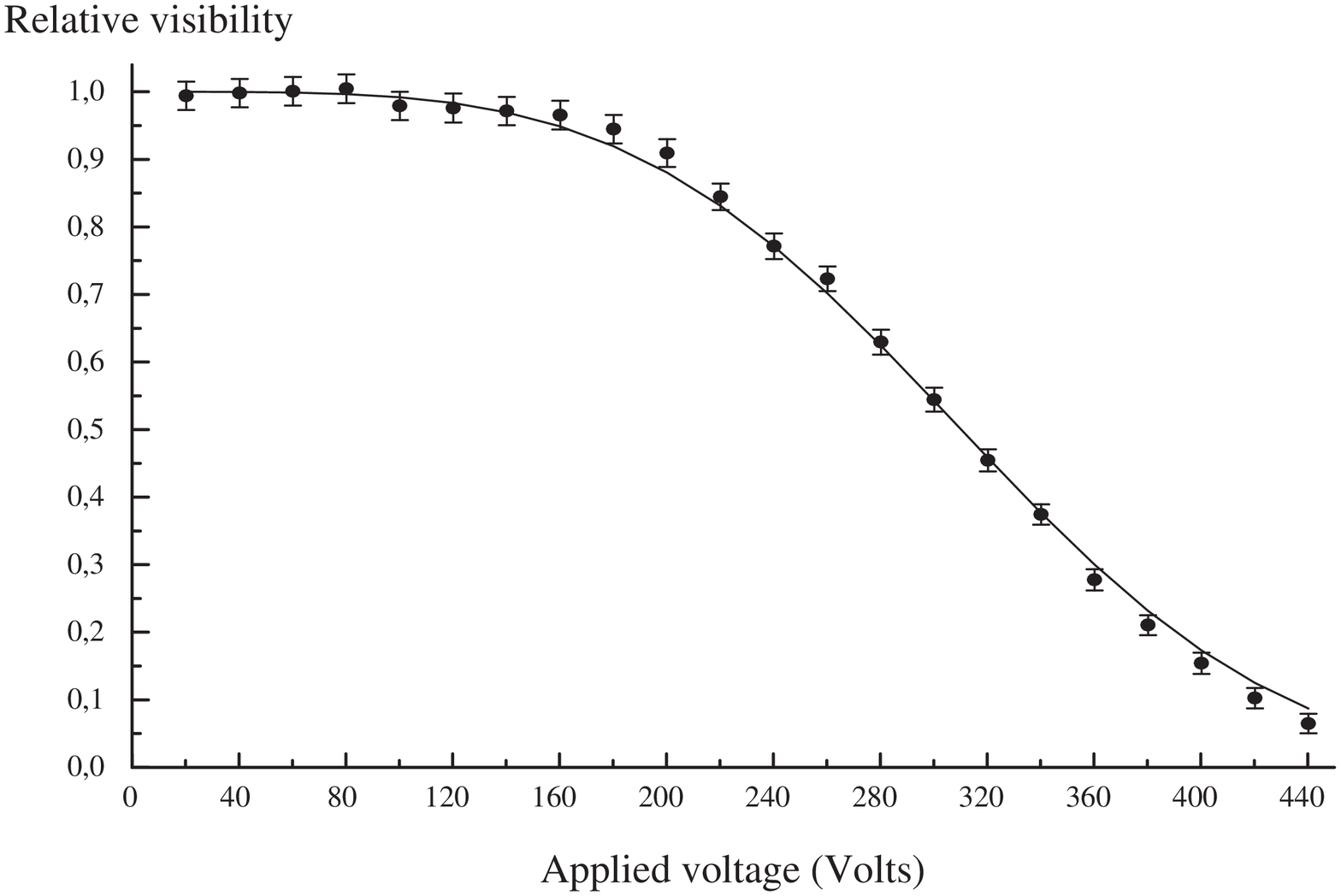}
\caption{\label{visibility} Relative fringe visibility
${\mathcal{V}}/{\mathcal{V}}_0$ with ${\mathcal{V}}_0 = 62$\% as a
function of the applied voltage $V_0$. The points are our
measurements and the full curve is our best fit using equations
(\ref{m4},\ref{m5}).}
\end{figure}

For each recording, the data points $I_i(n)$ have been fitted by a
function:

\begin{eqnarray}
\label{ex3} I_i(n) &=& I_{0i} \left[1 + {\mathcal{V}}_i\cos \psi_i
(n) \right] \nonumber \\
\mbox{with } \psi_i (n)&=& a_i + b_i n + c_i n^2
\end{eqnarray}

\noindent where $n$ labels the channel number, $a_i$ represents
the initial phase of the pattern, $b_i$ an ideal linear ramp and
$c_i$ the non-linearity of the piezo-drive. For the $V_0=0$
recordings, $a_i$, $b_i$ and $c_i$ have been adjusted as well as
the mean intensity $I_{0i}$ and the visibility ${\mathcal{V}}_i$.
For the $V \neq 0$ recording, we have fitted only $a_i$, $I_{0i}$
and ${\mathcal{V}}_i$, while fixing $b_i$ and $c_i$ to their value
$b_{i-1}$ and $c_{i-1}$ from the previous $V=0$ recording. We
think that our best phase measurements are given by the mean phase
$\bar{\psi}_i$ obtained by averaging $\psi_i (n)$ over the $471$
channels. The $1\sigma$ error bar of these mean phases are of the
order of $2-3$ mrad, increasing with the applied voltage up to
$23$ mrad because the visibility is considerably lower when the
applied voltage $V_0$ is large (see figure \ref{visibility}). This
rapid decrease of the visibility is due to the velocity dependence
of the phase and to the velocity distribution of the lithium
atoms.

\begin{figure}
\includegraphics[width = 8 cm,height= 6 cm,clip=true]{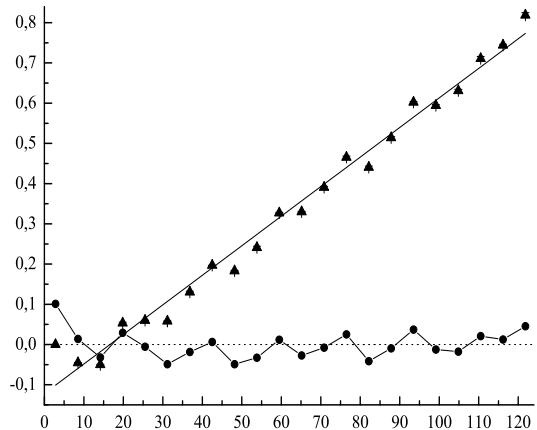}
\caption{\label{phasezero} The mean phase value $\left< \psi_i
\right>$, in radian, of the $V_0=0$ recordings is plotted as a
function of the recording starting time $t$ in minutes (with an
arbitrary origin). The straight line is the best linear fit,
corresponding to a phase drift of $ 7.5 \pm 0.2 $ mrad/minute. The
fit residuals are also plotted and from $t=15$ till $t=100$
minutes, the residuals exhibit an oscillating pattern with a
period close to $17$ minutes.}
\end{figure}

The mean phase values $\bar{\psi}_i$ values of the $V_0=0$
recordings are plotted in figure \ref{phasezero}: they present a
drift equal to $ 7.5 \pm 0.2 $ mrad/minute and some scatter around
this regular drift. The most natural explanation for this drift is
a change of the phase $\psi$ resulting from a variation of the
mirror positions $x_i$: $\psi$ changes by $1$ radian for a
variation of $(x_1+x_3-2x_2)$ equal to $53$ nm. We have verified
that the observed drift has the right order of magnitude to be due
to the differential thermal expansion of the structure supporting
the three mirrors: its temperature was steadily drifting at $1.17
\times 10^{-3}$ K/minute during the experiment and the support of
mirror $M_3$ differs from the other supports, as it includes a
piezo translation stage, which is replaced by aluminium alloy for
the mirrors $M_1$ and $M_2$. Presently, we have no explanation of
the phase scatter, which presents a quasi-periodic structure as a
function of time: its rms value is equal to $33$ milliradian and,
unfortunately, this scatter gives the dominant contribution to our
phase uncertainty.

\begin{figure}
\includegraphics[width = 8 cm,height= 7 cm,clip=true]{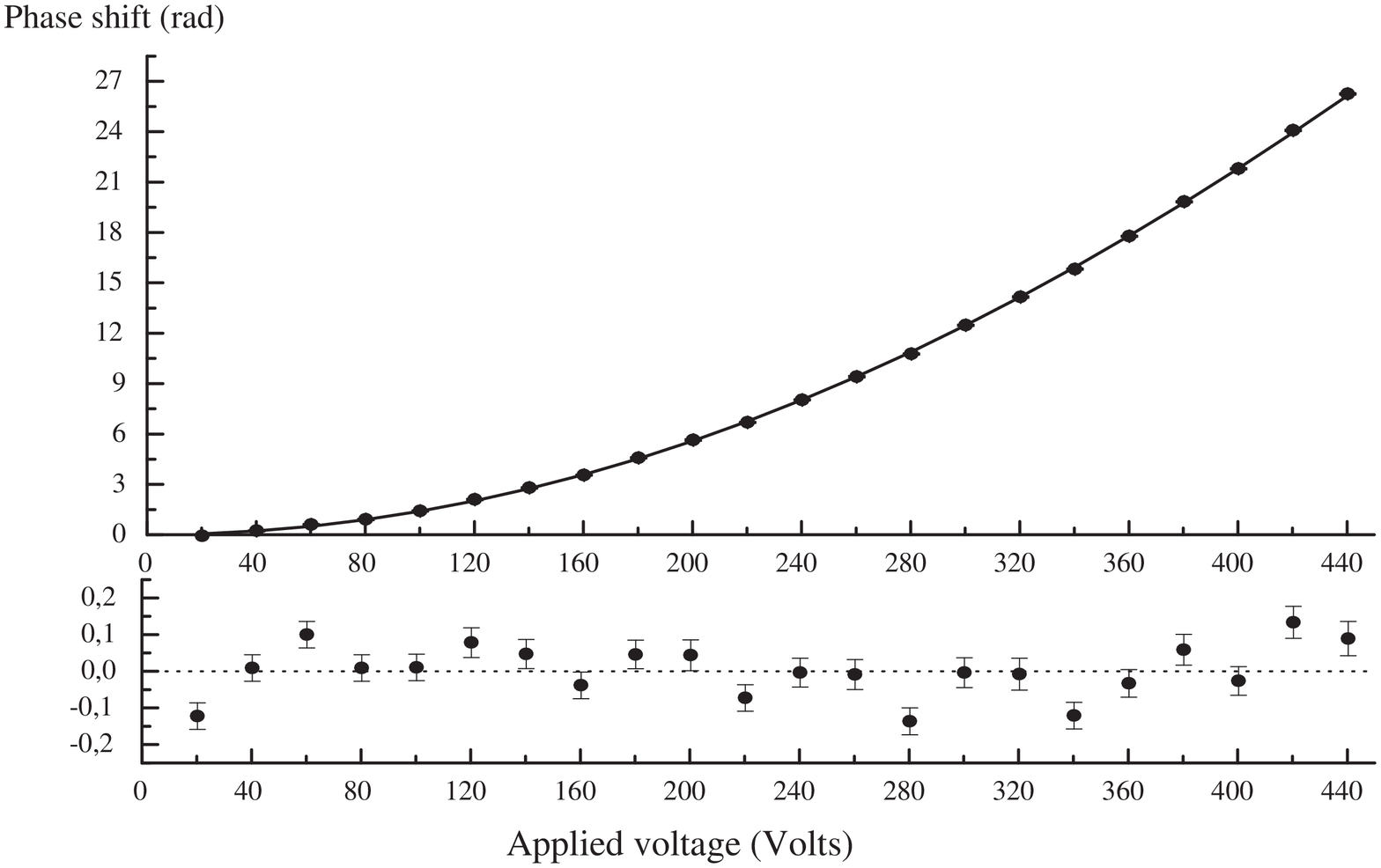}
\caption{\label{phaseshifts} The measured phase shift
$\left<\phi(V_0)\right>$ in radian is plotted as a function of the
applied voltage $V_0$: the best fit using equations
(\ref{n4},\ref{n5}) is represented by the full curve and the
residuals are plotted in the lower graph with an expanded scale.}
\end{figure}

The phase shift $\left<\phi(V_0)\right>$ due to the polarizability
effect (the average $\left<\right>$ recalls that our experiment
makes an average over the velocity distribution, as discussed
below) is taken equal to $\left<\phi(V_0)\right> = \bar{\psi}_i -
\left(\bar{\psi}_{i-1} + \bar{\psi}_{i+1}\right)/2 $ where the
recording $i$ corresponds to the applied voltage $V_0$: the
average of the mean phase of the two $V_0=0$ recordings done just
before and after is our best estimator of the mean phase of the
interference signal in zero field. In figure \ref{phaseshifts}, we
have plotted the phase shift $\phi(V_0)$ as a function of the
applied voltage $V_0$. We have chosen for the error bar on
$\phi(V_0)$ the quadratic sum of the $1\sigma$ error bar given by
the fit of $\left< \psi_i \right>$ and the $33$ milliradians rms
deviation of the $V_0=0$ phase measurements.

\section{Analysis of the signals: the effect of the lithium
velocity distribution}

To interpret the experimental data, we must take into account the
velocity distribution of the lithium atoms.

\subsection{Velocity averaging of the interference signal}

We assume that this velocity distribution is given by:

\begin{equation}
\label{m4} P(v)  = \frac{S_{\|}}{u \sqrt{\pi}} \exp\left[-\left(
(v-u)S_{\|}/u\right)^2\right]
\end{equation}
\noindent where $u$ is the most probable velocity and $S_{\|}$ is
the parallel speed ratio. With respect to the usual form of the
velocity distribution for supersonic beams, we have omitted a
$v^3$ factor which is traditionally introduced \cite{haberland85}
but when the parallel speed ratio $S_{\|}$ is large enough, this
$v^3$ has little effect and the main consequence of its omission
is a slight modification of the values of $u$ and $S_{\|}$. Here,
what we consider is the atoms contributing to the interferometer
signal and their velocity distribution may differ slightly from
the velocity of the incident beam, as Bragg diffraction is
velocity selective. The experimental signal can be written:

\begin{eqnarray}
\label{m5} I &=& I_0 \int dv P(v)  \left[1 + {\mathcal{V}}_0
\cos\left( \psi + \phi_m  \frac{u}{v} \right) \right] \nonumber \\
&=& I_0  \left[1 + \left<{\mathcal{V}}\right> \cos\left( \psi +
\left<\phi\right> \right) \right]
\end{eqnarray}

\noindent \noindent where $\phi_m $ is the value of the phase
$\phi$ for the velocity $v=u$. If we introduce $\delta = (v-u)/u$
and expand $(u/v)$ in powers of $\delta$ up to second order, the
integral can be taken exactly, as discussed in appendix B.
However, the accuracy of this approximation is not good enough
when $\phi_m$ is large and we have used direct numerical
integration to fit our data.

\subsection{Numerical fit of the data}

Using equations (\ref{m4}) and (\ref{m5}), we have fitted the
measured phase $\phi$ and visibility ${\mathcal{V}}$ as a function
of the applied voltage $V$. The phase measurements received a
weight inversely proportional to the square of their estimated
uncertainty and we have adjusted two parameters, the value of
$\phi_m/ V^2$ and the parallel speed ratio $S_{\|}$. The results
of the fits are presented in figures \ref{phaseshifts} and
\ref{visibility}. The agreement is excellent, in particular for
the phase data, and we deduce a very accurate value of $\phi_m/
V_0^2$:

\begin{equation}
\label{m71} \phi_m/ V_0^2 = (1.3870\pm 0.0010) \times 10^{-4}
\mbox{ rad/V}^2
\end{equation}

\noindent where the error bar is equal to $1\sigma$. The relative
uncertainty on $ \phi_m/ V_0^2$ is very small, $0.072$\%, which
proves the quality of our phase measurements. We also get an
accurate determination of the parallel speed ratio:

\begin{equation}
\label{m72} S_{\|} =  8.00 \pm 0.06
\end{equation}

\noindent This value of the parallel speed ratio is larger than
the predicted value for our lithium beam, $S_{\|, Li} = 6.2$ (see
above) and this difference can be explained by the velocity
selective character of Bragg diffraction.

\section{The electric polarizability of lithium}

The lithium electric polarizability is related to the value of
$\phi_m/ V_0^2$ by:

\begin{equation}
\label{p1} \frac{\phi_m}{V_0^2} = \frac{2 \pi \epsilon_0 \alpha
}{u} \times
 \left[\frac{2a}{h_0^2} - \frac{2h_0}{\pi}\right] \left[1 +
\frac{h_1^2}{h_0^2}\right] \end{equation}

\noindent All the geometrical parameters $a$, $h_0$ and $h_1$
describing the capacitor are known with a good accuracy and we
deduce a still very accurate value of the ratio of the electric
polarizability $\alpha$ divided by the mean atom velocity $u$:

\begin{equation}
\label{p2} \alpha/u = (2.283 \pm 0.008 ) \times 10^{-32} \mbox{
m}^2\mbox{.s}
\end{equation}

\subsection{Measurement of the mean atom velocity}

We have measured the mean atom velocity $u$ by various techniques

We have made measurements by Doppler effect, measured either on
the laser induced fluorescence signals or on the intensity of the
atomic beam which is reduced by atomic deflection due to photon
recoil. In the first case, the laser was making an angle close to
$49^{\circ}$ with the atomic beam and it is difficult to measure
this angle with sufficient accuracy.

\begin{figure}[h]
\includegraphics[width = 8 cm,height= 7 cm]{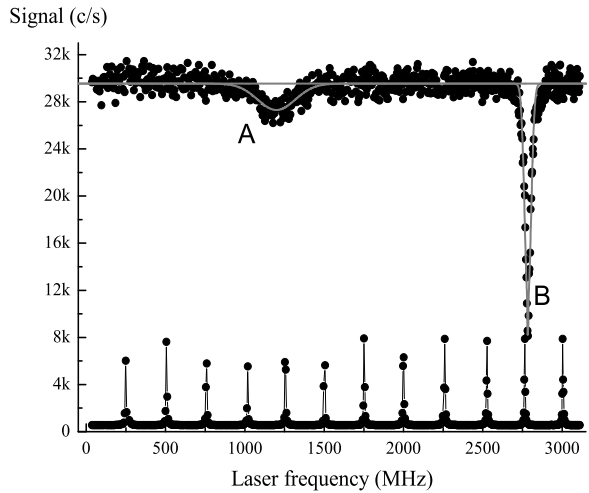}
\caption{\label{atomrecoil} Intensity losses recorded as a
function of the laser frequency in MHz. The peak A corresponds to
a laser beam almost contra-propagating with the atoms while the
peak corresponds to laser beam perpendicular to the atomic beam.
The lower curve is the transmission peaks of a Fabry-Perot
confocal interferometer used for frequency calibration.}
\end{figure}

In the second case, we have used a laser beam almost
contra-propagating with the atoms, so that the uncertainty on the
cosine of the angle is negligible. The signal appears as an
intensity loss on the atomic beam and, the loss is not very large
because we have used only one laser, so that the atoms are rapidly
pumped in the other hyperfine state. The experimental signal is
shown in figure \ref{atomrecoil}. From a fit of this data, we get
a value of the mean velocity $u = 1066.4 \pm 8.0$ m/s.

\begin{figure}
\includegraphics[width = 8 cm,height= 7 cm]{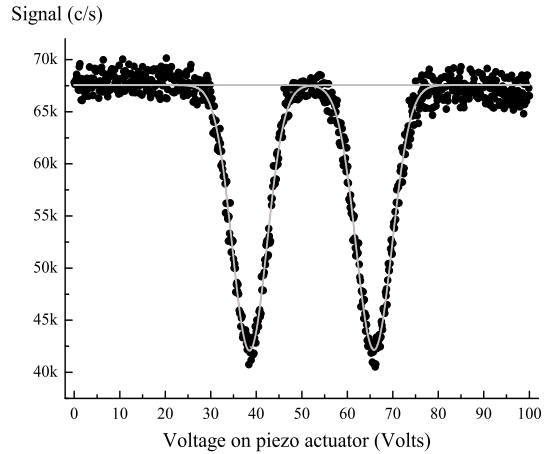}
\caption{\label{diffraction} Plot of the intensity of the atomic
beam as a function of the voltage applied on the piezoelectric
actuator inducing the rotation angle of mirror $M_2$. When the
Bragg condition is fulfilled, the direct beam intensity is reduced
by diffraction in the diffraction orders $\pm 1$. The dots
represent the measured intensity and the curves their best fit. In
a separate experiment, we have calibrated the rotation angle as a
function of the piezo voltage, $5.83\pm0.03$ $\mu$rad/V.}
\end{figure}

We have also recorded the diffraction probability as a function of
the Bragg angle, by tilting the mirror forming a standing wave.
This experiment is similar to the one described inour paper
\cite{miffre05} (see figure 3) but it is made with a lower power
density so that only the first order diffraction appears. The
diffraction is detected by measuring the intensity of the
zero-order atomic beam, as shown in figure \ref{diffraction}.
Using an independent calibration of the mirror rotation as a
function of the applied voltage on the piezo-actuator, we get a
measurement of the Bragg angle $ \theta_B = h/(mu \lambda_L) =
79.62 \pm 0.63 $ $\mu$rad corresponding to $u= 1065.0 \pm 8.4$
m/s.

These two values are very coherent and we may combine them to give
our best estimate of the most probable velocity $u$:

\begin{equation}
\label{p3} u = (1065.7 \pm 5.8 ) \mbox{m/s}
\end{equation}

We can compare this measurement with the theoretical prediction
for supersonic expansions. For a pure argon beam, in the limit of
an infinite parallel speed ratio, theory predicts $u = \sqrt{5
k_BT_0/m}$ m/s where $T_0$ is the nozzle temperature and $m$ the
argon atomic mass. With $T_0= 1073 \pm 11$ K, we get $u= 1056.7
\pm 5.5 $ m/s. Three small corrections must be made. We must
correct for the finite value of the argon parallel speed ratio
$S_{\|Ar}$, estimated to be $S_{\|Ar}=8.3$, using the
semi-empirical relation of Beijerinck and Verster
\cite{beijerinck81} and the associated correction reduces the most
probable velocity $u$ by a fraction equal to $0.75/S_{\|Ar}^2
\approx 1.09$\% (this correction is calculated in the limit of a
vanishing perpendicular temperature \cite{toennies77}). We must
replace the argon atomic mass by a weighted mean of the lithium
and argon atomic masses and, with $0.86$ mbar of lithium in $167$
mbar of argon, this correction increases the velocity by $0.21$\%.
Finally, we must take into account the velocity slip effect: the
light lithium atoms go slightly faster than the argon atoms. This
difference has been calculated by numerical simulation by P. A.
Skovorodko \cite{skovorodko04a} and this quantity is expected to
scale like $S_{\|Ar}^2$, so that the correction in our case is
estimated to be $2.42$\%. We thus predict a most probable velocity
$u = 1073.0\pm 5.6$ m/s, where the uncertainty comes solely from
the uncertainty on the temperature $T_0$. This value is in
satisfactory agreement with our measurements.

\subsection{The electric polarizability of lithium}

Using the measured value of the most probable velocity $u$, we get
the lithium electric polarizability of $^7$Li:

\begin{eqnarray}
\label{p4} \alpha &=&(24.33\pm 0.16)\times10^{-30} {\mbox{ m}}^3
\nonumber \\ &=& 164.19\pm 1.08 \mbox{ atomic units}
\end{eqnarray}

\noindent The final uncertainty bar is equal to $0.66$\%,
resulting from the quadratic sum of the $0.54$\% uncertainty on
the most probable velocity $u$, the $0.2$\% uncertainty on the
effective length of the capacitor, the $2\times 0.15 $\%
uncertainty due to the capacitor spacing $h_0$ and the $0.07$\%
uncertainty on the interferometric measurement. Unexpectedly, the
atom interferometry result has the smallest uncertainty!

Our measurement is a mean of the polarizability of the two
hyperfine sublevels $F=1$ and $F=2$ of $^7$Li. These two levels
have not exactly the same polarizability. The difference
$\Delta\alpha = \alpha (F=2, M_F=0) - \alpha(F=1,M_F =0)$ has been
measured with great accuracy by Mowat \cite{mowat72}. We can
express this result as a fraction of the mean polarisability,
$\Delta\alpha = -(3.0 \pm 0.1)\times10^{-6} \alpha $. This
difference is fully negligible with our present accuracy.

Our result is in excellent agreement with those of the previous
measurements of $\alpha$ which are considered as being reliable.
The 1934 measurement of Scheffers and Stark \cite{scheffers34},
which gave $\alpha = (12 \pm 0.6)\times10^{-30}$ m$^3$, is
generally considered to be incorrect. The first reliable
measurement is due to Bederson and co-workers \cite{salop61} in
1961, who obtained $\alpha = (20. \pm 3.0)\times10^{-30}$ m$^3$,
by using the E-H gradient balance method. In 1963, Chamberlain and
Zorn \cite{chamberlain63} obtained $\alpha = (22. \pm
2.)\times10^{-30}$ m$^3$ by measuring the deflection of an atomic
beam. Finally, in 1974, a second experiment was done by Bederson
and co-workers \cite{molof74}, using the same E-H gradient balance
method improved by the calibration on the polarizability of helium
atom in the $^3S_1$ metastable state, and they obtained the value
$\alpha = (24.3 \pm 0.5)\times10^{-30}$ m$^3$.

We may also compare our measurement with theoretical results. Many
calculations of $\alpha$ have been published and one can find a
very complete review with 35 quotations in table 17 of the paper
of King \cite{king97} published in 1997. Here is a very brief
discussion of the most important results:

\begin{itemize}
\item the first calculation of $\alpha$ was made in 1959 by
Dalgarno and Kingston \cite{dalgarno59} and gave $\alpha = 165$
a.u.. This calculation was rapidly followed by several other
works.

\item if we forget Hartree-Fock results near $170$ a.u., a large
majority of the published values are in the $162-166$ a.u. range.

\item in 1994, Kassimi and Thakkar \cite{takkar94} have made a
detailed study with two important results. They obtained a fully
converged Hartree-Fock value, $\alpha = 169.946$ a.u. and this
result, far from the experimental values, proves the importance of
electron correlation. They also made a series of nth-order
M\"oller-Plesset calculations with $n=2$, $3$ and $4$, from which
they extract their best estimate with an error bar, $\alpha =
164.2 \pm 0.1 $ a.u..

\item In 1996, Yan et al. \cite{yan96} have made an Hylleraas
calculation, with the final value $\alpha = 164.111 \pm 0.002$
a.u., this  value and its error bar resulting from a convergence
study.

\end{itemize}

Our result is extremely close to these two very accurate
calculations of $\alpha$. Two minor effects have not been taken
into account in these calculations, namely relativistic correction
and finite nuclear mass correction, but these two effects are
quite small.

The relativistic correction on the polarizability has been studied
by Lim et al. \cite{lim99}. They have made different calculations
(Hartree-Fock, second order M\"oller-Plesset, coupled cluster CCSD
and CCSD(T)) and in all cases, their relativistic result is lower
than the non-relativistic result with a difference in the
$0.05-0.07$ a.u. range. We do not quote here their results: even
the Hartree-Fock value is lower by $0.45$ a.u. than the one of
reference \cite{takkar94}, because the chosen basis set is too
small.

As far as we know, no calculation of $\alpha$ has been made taking
into account the finite nuclear mass. An order of magnitude of the
associated correction should be given by the hydrogenic
approximation: $\alpha \propto \left[1+(m/M)\right]^3$ and in this
approximation, the polarisability of $^7$Li should be larger than
its $M= \infty $ value by $0.04$ a.u. but one should not expect
this approximation to predict even the sign of the correction. A
high accuracy calculation of the finite mass effect is surely
feasible, following the Hylleraas calculations of Yan and Drake,
who have already evaluated the finite mass effect on some energies
\cite{yan95a} and oscillator strengths \cite{yan95b} of lithium
atom.

\section{Conclusion}

We have made a measurement of the electric polarizability of
lithium atom $^7$Li by atom interferometry and we have obtained
$\alpha = (24.33\pm 0.16)\times10^{-30} {\mbox{ m}}^3 = 164.19\pm
1.08$ a.u. with a $0.66$\% uncertainty. Our measurement is in
excellent agreement with the most accurate experimental value
obtained by Bederson and coworkers \cite{molof74} in 1974 and we
have reduced the uncertainty by a factor three. Our result is also
in excellent agreement with the best theoretical estimates of this
quantity due to Kassimi and Thakkar \cite{takkar94} and to Yan et
al. \cite{yan96}. The neglected corrections (relativistic effect,
finite nuclear mass effect) should be at least ten times smaller
than our present error bar.

Our measurement is the second measurement of an electric
polarizability by atom interferometry, the previous experiment
being done on sodium atom by D. Pritchard and co-workers
\cite{ekstrom95} in 1995 (see also \cite{schmiedmayer97} and
\cite{roberts04}). This long delay is explained by the difficulty
of running an atom interferometer with spatially separated beams.
Using a similar experiment, J. P. Toennies and co-workers have
compared the polarizabilities of helium atom and helium dimer but
this work is still unpublished \cite{toennies03}.

We want now to insist on the improvements we have done with
respect to the other measurement of an electric polarizability by
atom interferometry, due to D. Pritchard and co-workers
\cite{ekstrom95} :
\begin{itemize}
\item The design of our capacitor permits an analytical
calculation of the $E^2$ integral along the atomic path. This
property is important for a better understanding of the influence
of small geometrical defects of the real capacitor. In the present
experiment, the uncertainty on the $E^2$ integral is equal to
$0.36$\% and we think that it is possible to reduce this
uncertainty near $0.1$ \% with an improved construction.

\item We have obtained a very good phase sensitivity of our atom
interferometer: from our recordings, we estimate this phase
sensitivity near $16$ mrad/$\sqrt{\mbox{Hz}}$. The accuracy
achieved on phase measurement has been limited by the lack of
reproducibility of the phase between consecutive recordings. We
will stabilize the temperature of the rail supporting the three
mirrors, hoping thus to improve the phase stability. Even if we
have not been able to fully use our phase sensitivity, we have
obtained set of phase-shifts measurements exhibiting an excellent
consistency and accuracy, as shown by the quality of the fit of
figure \ref{phaseshifts} and by the accuracy, $\pm 0.072$\%, of
the measurement of the quantity $\phi_m/V_0^2$.

\item In his thesis \cite{roberts02a}, T. D. Roberts  reanalyzes
the measurement of the electric polarizability of sodium atom made
by C. R. Ekstrom et al. \cite{ekstrom95}: he estimates that a weak
contribution of sodium dimers to the interference signals can be
present as material gratings diffract sodium dimers as well as
sodium atoms and he estimates that, in the worst case, this
molecular signal might have introduced a systematic error as large
as $2$\% on the sodium polarizability result. Our interferometer
is species selective thanks to the use of laser diffraction and
this type of error does not exist in our experiment. Only lithium
atoms are diffracted and even, with our choice of laser
wavelength, only the $^7$Li isotope contributes to the signal.

\end{itemize}

The main limitation on the present measurement of the electric
polarisability of lithium $^7$Li comes from the uncertainty on the
most probable atom velocity $u$. With some exceptions, like the
Aharonov-Casher phase shift \cite{aharonov84} which is independent
of the atom velocity, a phase-shift induced by a perturbation is
inversely proportional to the atom velocity, at least when a
static perturbation is applied on one interfering beam. This is a
fundamental property of atom interferometry and clever techniques
are needed to overcome this difficulty:

\begin{itemize}
\item T. D. Roberts et al. \cite{roberts04} have developed a way
of correcting the velocity dependence of the phase shift by adding
another phase shift with an opposite velocity dependence. They
were thus able to observe fringes with a good visibility up to
very large phase shift values.

\item our present results prove that a very accurate measurement
can be made in the presence of an important velocity dispersion
without any compensation of the associated phase dispersion, but
by taking into account the velocity distribution in the data
analysis.
\end{itemize}
In these two cases, one must know very accurately a velocity, the
most probable velocity in our case and the velocity for which the
correction phase cancels in the case of reference
\cite{roberts04}. The uncertainty on this velocity may finally be
the limiting factor for high precision measurements. Obviously
other techniques can be used to solve this difficulty.

Finally, we think that our experiment illustrates well two very
important properties of atom interferometry:

\begin{itemize}

\item The sensitivity of atom interferometry is a natural
consequence of the well-known sensitivity of interferometry in
general, which is further enhanced in the case of atom by the
extremely small value of the de Broglie wavelength. Our phase
measurement is in fact a direct mesaurement of the increase
$\Delta v$ of the atom velocity $v$ when entering the electric
field. $\Delta v$ is very simply related to the observed phase
shift:

\begin{equation}
\label{c1} \frac{\Delta v}{v} = \frac{\lambda_{dB}}{L_{eff}}
\times\frac{ \phi }{2 \pi}
\end{equation}

\noindent This variation is extremely small, with $\Delta v/v
\approx 4\times 10^{-9}$ for the largest electric field used in
this experiment, corresponding to $\phi \approx 25$ rad. Our
ultimate sensitivity corresponds to a phase $\phi \approx 3$
milliradian which means that we can detect a variation $\Delta v/v
\approx 6 \times 10^{-13}$, whereas the velocity distribution has
a FWHM width equal to $21$\%!

\item when an atom propagates in the capacitor placed in our atom
interferometer, its wavefunction samples two regions of space
separated by a distance $\sim 90$ $\mu$m with a macroscopic
object, the septum, lying in between and this situation extends
over $75$ microsecond duration, without inducing any loss of
coherence. This consequence of quantum mechanics remains very
fascinating!

\end{itemize}

\section{Acknowledgements}

We thank CNRS SPM and R\'egion Midi Pyr\'en\'ees for financial
support. We thank P. A. Skovorodko and J. L. Heully for helpful
information.

\section{Appendix A: detailed calculation of the capacitor electric
field}

The first step is to calculate the Fourier transform
$\tilde{V}(k)$ of $V(x=h,z)$ defined by equation (\ref{n4}):

\begin{eqnarray}
\label{a1} \tilde{V}(k)  & = & \frac{1}{\sqrt{2\pi}}
\int_{-a}^{+a}V_0 \exp\left( -ikz\right) dz \nonumber \\ & =&
\frac{2 V_0 }{\sqrt{2\pi}} \times \frac{\sin\left( ka\right)}{k}
\end{eqnarray}

\noindent The potential $V(x,z)$ given by equation (\ref{n5}) can
then be calculated and, from $V(x,z)$, we can deduce the electric
field everywhere and in particular, on the septum surface where it
is parallel to the $x$-axis:

\begin{equation}
\label{a2} E_{x}(x=0, z)   =  \frac{1}{\sqrt{2\pi}}
\int_{-\infty}^{+\infty} \tilde{V}(k) \tilde{g}(k)
 \exp\left(ikz\right) dk
\end{equation}

\noindent with $\tilde{g}(k) = k/ \sinh\left(k h\right)$. The
electric field $E_{x}(x=0,z)$ is given by the inverse Fourier
transform of the product of two functions $\tilde{V}(k)$ and
$\tilde{g}(k) $. Therefore, the field $E_{x}(x=0, z)$ is the
convolution of their inverse Fourier transforms which are $V(x= h,
z)$ and $g(z)$:

\begin{equation}
\label{a3} E_{x}(x=0, z)  = \int_{-\infty}^{+\infty} V(x= h, z_1)
g(z - z_1) dz_1
\end{equation}

\noindent Using reference \cite{gradshteyn80} (equation 6 of
paragraph 4.111, p. 511), we get an explicit form of $g(z)$:

\begin{equation}
\label{a4} g(z) = \frac{\pi^2}{2 h^2 \cosh^2\left(\frac{\pi z}{2
h}\right)}
\end{equation}

\noindent This result proves that the field decreases
asymptotically like $\exp\left( -\pi |z|/h \right)$, when
$\left|z\right| -a  \gg h $. We can also get a closed form
expression of the electric field  $E_{x}(x=0, z)$ but we may get
the integral of $E^2$ without this result, simply by using the
Parseval-Plancherel theorem and reference \cite{gradshteyn80}
(equation 4 of paragraph 3.986, p. 506):

\begin{equation}
\label{a51} \int_{-\infty}^{+\infty}E_{x}(x=0,z)^2 dz = \frac{2a
V_0^2}{h^2}\left[\coth\left(\frac{\pi a}{h}\right) - \frac{h}{\pi
a}\right]
\end{equation}

\noindent The atoms sample the electric field at a small distance
of the septum surface and we need the integral of $E^2$ along
their mean trajectory. This mean trajectory is not parallel to the
septum, but it is easier to calculate this integral along a
constant $x$ line. From Maxwell's equations, one gets the first
correction terms to the field when $x$ does not vanish:

\begin{eqnarray}
\label{a52} E_x(x, z) & \approx & E_x(x=0, z) -
\frac{x^2}{2}\times\frac{\partial^2 E_x}{\partial z^2} \nonumber \\
E_z(x,z) &\approx & x \frac{\partial E_x}{\partial z}
\end{eqnarray}
\noindent where the derivatives are calculated for $x=0$. After an
integration by parts, one gets:

\begin{eqnarray}
\label{a6} & & \int_{-\infty}^{+\infty}\left[E_{x}^2(x,z) +
E_{y}^2(x,z) \right] dz \nonumber \\
& = & \int_{-\infty}^{+\infty}\left[ E_{x}^2(x=0,z) + 2 x^2
\left[\frac{\partial E_x}{\partial z}\right]^2\right] dz
\end{eqnarray}

\noindent  where we have kept only the first non vanishing
correction term in $x^2$. The calculation of the integral is also
done with the Parseval-Plancherel theorem and, after some algebra,
we get:

\begin{eqnarray}
\label{a7} 2 x^2 \int_{-\infty}^{+\infty} \left[\frac{\partial
E_x}{\partial z}\right]^2 dz &=&x^2 \frac{V_0^2}{h^3}
\left[\frac{2}{3} - \frac{\frac{\pi a}{h}\coth\left(\frac{\pi
a}{h}
\right) - 1 }{\sinh^2\left(\frac{\pi a}{h}\right)}\right] \nonumber \\
& \approx & x^2 \frac{ 2 V_0^2}{3 h^3}
\end{eqnarray}

\noindent where the approximate result is obtained by neglecting
the exponentially small terms of the order of $\exp\left[ -2\pi
h/\left(a \right)\right]$.

\section{Appendix B: velocity average of the interference signals}

We want to calculate:

\begin{equation}
\label{B1} I = I_0 \int dv P(v)  \left[1 + {\mathcal{V}}_0
\cos\left( \psi + \phi_m \frac{u}{v} \right) \right]
\end{equation}

\noindent with the velocity distribution given by:

\begin{equation}
\label{B2} P(v)  = \frac{S_{\|}}{u \sqrt{\pi}} \exp\left[-\left(
(v-u)S_{\|}/u\right)^2\right]
\end{equation}
\noindent Noting $\delta = (v-u)/u$ and expanding $(u/v)$ in
powers of $\delta$ up to second order, the integral becomes:

\begin{eqnarray} \label{B3} I/I_0 & =&  \frac{S_{\|}}{u
\sqrt{\pi}} \int d\delta
\exp\left[-\delta^2 S_{\|}^2\right] \nonumber \\
&\times & \left[1 + {\mathcal{V}}_0 \cos\left[ \psi + \phi_m
\left(1-\delta +\delta^2\right) \right] \right]
\end{eqnarray}
\noindent which can be taken exactly:

\begin{eqnarray}
\label{B41} I / I_0 &= &\left[1 + \left<{\mathcal{V}}\right>
\cos\left( \psi + \left<\phi\right> \right)\right] \\
\label{B42} \left<{\mathcal{V}}\right> &= & {\mathcal{V}}_0
\frac{S_{\|}}{\left[S_{\|}^4 +\phi_m^2\right]^{1/4}}
\exp\left[-\frac{\phi_m^2 S_{\|}^2}{4\left( S_{\|}^4
+\phi_m^2\right)} \right] \\
\label{B43} \left<\phi \right> &=& \phi_m  +\frac{1}{2}
\arctan\left[\frac{\phi_m}{ S_{\|}^2} \right] - \frac{\phi_m^3
}{4\left( S_{\|}^4 +\phi_m^2\right)}
\end{eqnarray}

\begin{figure}
\includegraphics[width = 8 cm,height= 7 cm]{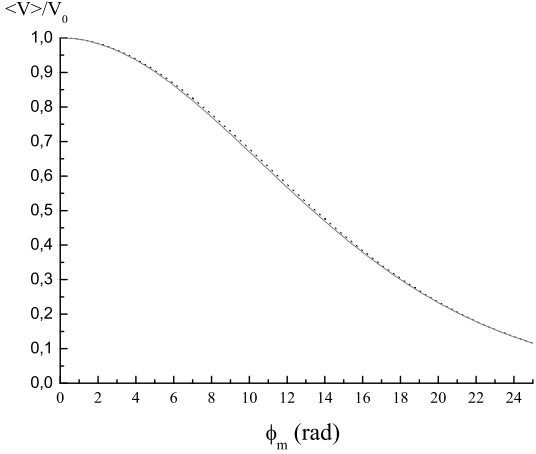}
\caption{\label{visibilityaverage} Calculation of the average
visibility $\left<{\mathcal{V}}\right>$ in the case $S_{\|} = 8$
as a function of $\phi_m$. The approximate result (equation
\ref{B42}) is plotted with a dashed line while the result of the
numerical integration is plotted by the full line.}
\end{figure}

We have tested this approximation by comparing this approximate
formula with the result of a computer program, for a parallel
speed ratio $S_{\|} = 8$, corresponding to our experimental case.
As shown on figure \ref{visibilityaverage}, the agreement is very
satisfactory, at least with our present accuracy on visibility
values, with differences of the order of $1$\%.

\begin{figure}
\includegraphics[width = 8 cm,height= 7 cm]{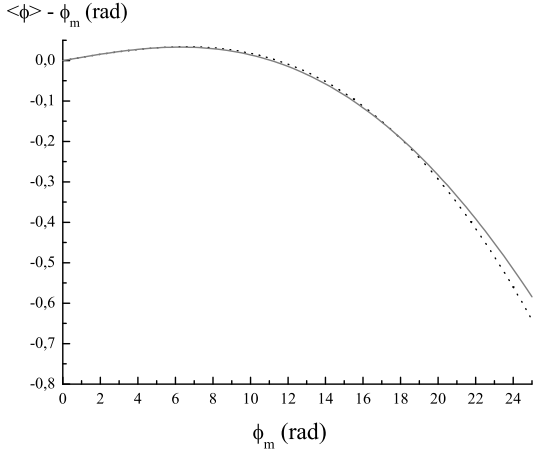}
\caption{\label{phaseaverage} Calculation of the average phase
$\left<\phi\right>$ in the case $S_{\|} = 8$. The quantity
$\left<\phi \right> - \phi_m $ is plotted as a function of
$\phi_m$: dashed line for the approximate result (equation
\ref{B43}) and full line for the result of the numerical
integration.}
\end{figure}

We have also studied the difference $\left(\left<\phi
\right>-\phi_m \right)$ as a function of $\phi_m$ and the results
are presented in figure \ref{phaseaverage}. This difference can
reach large values, for instance $-0.64$ rad when $\phi_m \approx
25$ rad. One may remark that, as obvious on equation (\ref{B43}),
the difference $\left(\left<\phi \right> -\phi_m \right)$ is
linear in $\phi_m$ and positive, when $\phi_m$ is small, and
becomes negative and roughly cubic in $\phi_m$ for larger $\phi_m$
values. If the parallel speed ratio $S_{\|}$ is large, $S_{\|}\gg
1$, as long as the linear term is dominant, the velocity averaged
phase is given by:

\begin{equation}
\label{B5} \left<\phi \right> = \phi_m \left[ 1 + \frac{1}{2
S_{\|}^2} \right]
\end{equation}

\noindent and not by $\phi_m$. With $S_{\|} = 8$, the approximate
and numerical results are almost equal as long as $\phi_m < 18$,
but their difference increases rapidly for larger $\phi_m$ values,
being close to $ 0.05$ radian when $\phi_m = 25$: even if this
difference is a very small fraction of $\phi_m$, this difference
is not fully negligible and we have decided not to use the
approximate analytical results (\ref{B42}) and (\ref{B43}) to fit
the data.

\begin{thebibliography}{}

\bibitem{bonin97} K. D. Bonin and V. V. Kresin, Electric-Dipole
Polarizabilities of Atoms, Molecules and Clusters (World
Scientific, 1997)

\bibitem{molof74} R. W. Molof, H. L. Schwartz, T. M. Miller and
B. Bederson, Phys. Rev. A {\bf 10}, 1131 (1974)

\bibitem{amini03} J. M. Amini and H. Gould, Phys. Rev. Lett. {\bf
91}, 153001 (2003)

\bibitem{ekstrom95} C. R. Ekstrom, J. Schmiedmayer, M. S. Chapman,
T. D. Hammond and D. E. Pritchard, Phys. Rev. A {\bf 51}, 3883
(1995)

\bibitem{delhuille02a} R. Delhuille, C. Champenois, M. B\"uchner, L.
Jozefowski, C. Rizzo, G. Tr\'enec and J. Vigu\'e, Appl. Phys. {\bf
B 74}, 489 (2002) 7

\bibitem{miffre05} A. Miffre, M. Jacquey, M. B\"uchner, G. Tr\'enec
and J. Vigu\'e, Eur. Phys. J. D {\bf 33}, 99 (2005)

\bibitem{miffre05b} A. Miffre, M. Jacquey, M. B\"uchner, G. Tr\'enec
and J. Vigu\'e, J. Chem. Phys. 122, 094308 (2005)

\bibitem{miffre05a} A. Miffre, M. Jacquey, M. B\"uchner, G. Tr\'enec
and J. Vigu\'e, submitted to Phys. Rev. A; preprint available on
https://hal.ccsd.cnrs.fr/ccsd-00005359

\bibitem{roberts04} T. D. Roberts, A. D. Cronin, M. V. Tiberg and
D. E. Pritchard, Phys. Rev. Lett. {\bf 92}, 060405 (2004)

\bibitem{shimizu92} F. Shimizu, K. Shimizu and H. Takuma, Jpn. J.
Appl. Phys. {\bf 31}, L436 (1992)

\bibitem{nowak98} S. Nowak, N. Stuhler, T. Pfau and J. Mlynek, Phys. Rev.
Lett. {\bf 81}, 5792 (1998)

\bibitem{nowak99} S. Nowak, N. Stuhler, T. Pfau and J. Mlynek,
Appl. Phys. B {\bf 69}, 269 (1999)

\bibitem{rieger93} V. Rieger, K. Sengstock, U. Sterr, J. H. M\"uller
and W. Ertmer, Opt. Comm. {\bf 99}, 172 (1993)

\bibitem{morinaga96} A. Morinaga, N. Nakamura, T. Kurosu and N. Ito,
Phys. Rev. A {\bf 54}, R21 (1996)

\bibitem{sangster93} K. Sangster, E. A. Hinds, S. M. Barnett and E.
Riis, Phys. Rev. Lett. {\bf 71}, 3641 (1993)

\bibitem{sangster95} K. Sangster, E. A. Hinds, S. M. Barnett, E.
Riis and A. G. Sinclair, Phys. Rev. A {\bf 51}, 1776 (1995)

\bibitem{zeiske95} K. Zeiske, G. Zinner, F. Riehle and J. Helmcke,
Appl. Phys. B {\bf 60}, 295 (1995)

\bibitem{aharonov84} Y. Aharonov and A. Casher, Phys. Rev. Lett.
{\bf 53}, 319 (1984)

\bibitem{cheval} Laser Cheval, website: http://www.cheval-freres.fr

\bibitem{keith91} D. W. Keith, C. R. Ekstrom, Q. A. Turchette and D.
E. Pritchard, Phys. Rev. Lett. {\bf 66}, 2693 (1991)

\bibitem{schmiedmayer97} J. Schmiedmayer, M. S. Chapman, C. R.
Ekstrom, T. D. Hammond, D. A. Kokorowski, A. Lenef, R. A.
Rubinstein, E. T. Smith and D. E. Pritchard, in Atom
interferometry edited by P. R. Berman (Academic Press 1997), p 1

\bibitem{giltner95b} D.M. Giltner, R. W. McGowan and Siu Au Lee,
Phys. Rev. Lett. {\bf 75}, 2638 (1995)


\bibitem{beijerinck81} H. C. W. Beijerinck and N. F. Verster,
Physica {\bf 111C}, 327 (1981)

\bibitem{toennies77} J. P. Toennies and K. Winkelmann, J.
Chem. Phys. {\bf 66}, 3965 (1977)

\bibitem{haberland85} H. Haberland, U. Buck and M. Tolle, Rev.
Sci. Instrum. {\bf 56}, 1712 (1985)

\bibitem{skovorodko04a} P. A. Skovorodko, 24th International
Symposium on Rarefied Gas Dynamics, AIP Conference Proceedings
{\bf 762}, 857 (2005) and private communication.

\bibitem{mowat72} J. R. Mowat, Phys. Rev. A {\bf 5}, 1059 (1972)

\bibitem{scheffers34} H. Scheffers and J. Stark, Phys. Z. {\bf 35}, 625
(1934)

\bibitem{salop61} A. Salop, E. Pollack and B. Bederson, Phys. Rev.
{\bf 124}, 1431 (1961)

\bibitem{chamberlain63} G. E. Chamberlain and J. C. Zorn, Phys.
Rev. {\bf 129}, 677 (1963)

\bibitem{king97} F. W. King, J. Mol. Structure (Theochem) {\bf
400}, 7 (1997)

\bibitem{dalgarno59} A. Dalgarno and A. E. Kingston, Proc. Roy.
Soc. {\bf 73}, 455 (1959)

\bibitem{takkar94} N. E. Kassimi and A. J. Thakkar, Phys. Rev. A
{\bf 50}, 2948 (1994)

\bibitem{yan96} Z. C. Yan, J. F. Babb, A. Dalgarno and G. W. F.
Drake, Phys. Rev. A {\bf 54}, 2824 (1996)

\bibitem{lim99} I. S. Lim, M. Pernpointer, M. Seth, J. K.
Laerdahl, P. Schwerdtfeger, P. Neogrady and M. Urban, Phys. Rev. A
{\bf 60} 2822 (1999)

\bibitem{yan95a} Z. C. Yan and G. W. F. Drake, Phys.
Rev. A {\bf 52}, 3711 (1995)

\bibitem{yan95b} Z. C. Yan and G. W. F. Drake, Phys. Rev. A
{\bf 52}, R4316 (1995)

\bibitem{roberts02a} T. D. Roberts, Ph. D. thesis (unpublished), MIT
(2002)

\bibitem{toennies03} J. P. Toennies, private communication (2003)

\bibitem{gradshteyn80} I. S. Gradshteyn and I. M. Ryzhik, Tables
of integrals, series and products, 4th edition, Academic Press
(1980)


\end{thebibliography}
\end{document}